# Synthesis of MgB$_2$ from elements

**N. N. Kolesnikov[*], M. P. Kulakov,**

Institute of Solid State Physics of Russian Academy of Sciences, Chernogolovka, Moscow District, 142432 Russia

**Abstract**

Superconducting at 40 K MgB$_2$ -samples were obtained by direct reaction from elements in molybdenum crucibles under argon pressure. Pressure allows to provide annealing at temperature up to 1400$^o$C, that resulting in rise of T$_c$ and compactness of the ceramics, suggesting that there is a homogeneity range of composition for the compound.

## 1. Introduction

Recent findings of superconductivity (SC) in MgB$_2$ [1,2] gave interest to its synthesis, doping, and phase formation and equilibrium in the system. There were reports on MgB$_2$ synthesis in sealed Ta-tubes [2], and on sintering in Ta–foils [3] with slightly different results in properties of obtained ceramics. The first attempts to sputter MgB$_2$ films [4] gave only 12 K for the SC-transition, and that contradicts to the existing phase T-x diagram of Mg-B system [5].

We report here a method of MgB$_2$ synthesis in Mo-crucibles under argon pressure, and give some properties of obtained ceramics, which support an idea of homogeneity range existing for MgB$_2$ composition.

## 2. Experiment

For the synthesis, we used amorphous boron powder and lump metal magnesium, both with purity better than 99,95%. Boron powder was pelletized under the load about 2 ton/cm$^2$ and placed at the bottom of Mo-crucible. Lump magnesium was laid up on the B-pellets, and the crucible (Fig. 1) was closed with a threaded cap. The inner crucible diameter was of 12.5 mm in most cases, but it could be enlarged up to 50 mm with the same wall thickness of 3 mm. The crucible was placed into the medium-pressure furnace with a resistive heater. Argon pressure is necessary at the synthesis because of Mg-vapor pressure being equal to 1 bar at 1100$^o$C and about 8 bar at 1400$^o$C, so magnesium loss during synthesis is inevitable in non-sealed systems. The furnace was preliminarily pumped out down to 10$^{-5}$ bar for 2 hours and then filled with Ar up to 10-12 bar. Actual Ar pressure was ~20 bar at 1000$^o$ and 27 bar at 1400$^o$C.

On heating, Mg melts at 650$^o$C, and reaction with Boron starts around 900$^o$C, that giving rise to the local in-crucible temperature and partial magnesium vaporization. In result, Mg-melt fills the cap threads and encapsulates the charge completely to react with boron, forming MgB$_2$. We found that the load more than 2 ton/cm$^2$ at pelletizing was ineffective, so as in this case dense ceramics at the reaction front blocked the Mg-supply to the front, and the reaction stopped, leaving the core parts of the boron charge untouched. In all other cases, the synthesized ceramics of MgB$_2$ sagged on to the bottom of the crucible as a dense cylinder which could be extracted afterwards, so as magnesium melt did not react with molybdenum. The heating could be continued up to 1400$^o$C with results, which are described in the next section.

Powder X-ray analysis of obtained ceramic samples was made with the Siemens D-500 diffractometer in the interval 20$^o$ < 2θ < 80$^o$ with steps of 0.02$^o$, Cu-Kα1 radiation being applied.

SC-transition temperature was determined by registering AC-susceptibility in the temperature interval of 4.2 to 100 K. Also the standard 4-probe method was used to obtain the resistance dependence on temperature.

The density of samples was measured by hydrostatic weighing in toluene at 25$^o$C.

---

[*] Corresponding author: fax: 7-(096)-52-49701, E-mail: nkolesn@issp.ac.ru

## 3. Results and discussion

The syntheses gave us compact ceramic cylinders (Fig. 2) with volume of about 30% of inner crucible volume. The output could be larger, if we had used crystalline boron for the starting material. The measured density was 2.42 g/cm$^3$ for syntheses at 1000$^o$C with subsequent rapid cooling, and only 2.23 g/cm$^3$ after MgB$_2$ heating up to 1400$^o$C and then keeping this temperature for an hour. This drop in density was due to Mg- evaporation and formation of numerous small voids (5-30 microns) in the ceramics.

The powder X-ray diffraction pattern (Fig. 3) shows, besides MgB$_2$, the presence of small amounts of elemental magnesium and magnesium oxide MgO. The elemental Mg comes from the condensation of equilibrium Mg-gas phase on cooling, whereas MgO, as we believe, is due to Mg-gettering of oxygen adsorbed on boron powder. In the case of 1400$^o$-annealing, along with Mg-evaporation the formation of MgB$_4$ is registered by XRD-data on occurrence of two the most intensive reflections of the phase.

The SC-properties of the obtained MgB$_2$ samples also reflect to some extent the synthesis conditions. The ceramics sintered at 1000$^o$C reveals the SC transition at 36 K by AC-susceptibility (Fig. 4), whereas the ceramics annealed at 1400$^o$C shows the SC transition at 38 K by AC-susceptibility and 40 K by resistance (Fig. 4, 5). We believe this trend in transition temperature to originate from the existence of the homogeneity range in the composition of MgB$_2$. The more the composition is shifted to B-side, the more high $T_c$ the phase reveals. This conclusion is in compliance with very low $T_c \sim 12$ K for Mg-enriched films [4], and with higher $T_c$ for 1400$^o$-annealed ceramics in our experiments. It is worth noting that doping MgB$_2$ with copper at 1000$^o$C, as in [3], gives actually the same result in $T_c$ as the high-temperature annealing (Fig. 4). Evidently, the intentional large-scale Cu-doping of MgB$_2$ gives only the phase composition shift to the B-side with possible appropriate rise of $T_c$. The repetition of the experiment, made in [3], in our conditions resulted in accumulation of Cu-rich phase MgCu$_2$ just before the front of MgB$_2$ formation, because the first one is low-melting (~819$^o$C) and insoluble in the second.

Both the known T-x diagram of the Mg-B system [5] and its computational elucidation [6] do not suggest such a homogeneity range. We also are unable to make difference between two borders of supposed range (one with excess of Mg and the other with co-existing MgB$_4$) using our XRD-data: both give the same result $a_o=3.086(4)$, $c_o=3.520(0)$. It is also doubtful that the difference will be found by micro-probe analysis, so as Mg-counts drop only ~ 35% when the probe moves from MgB$_2$ to the pure boron phase [7]. Nevertheless, the ceramics from the sides look externally very much unlike: the one synthesized at 1000$^o$C is brownish-black and somewhat loose for touch, whereas the other one annealed at 1400$^o$C has light-bronze color and is very tough to break. As considered in [8], that reflects very different carrier in such samples. The study needs to be continued.

## 4. Conclusion

It is shown that MgB$_2$ ceramics is proven to be produced on relatively large scale in Mo-crucible under moderate argon pressure. Ceramics synthesized at 1000$^o$C seems black and loose, and has slightly lower $T_c$, than the one annealed at 1400$^o$C which is very durable mechanically. As yet, there are no other parameters to differentiate these two ceramics.

## Acknowledgments

We are grateful to Prof. V.V. Ryazanov for providing us with resistivity and susceptibility measurements, and to I. I. Zverkova for help with X-ray analysis.

**Figure captions**

1. Molybdenum crucible (cross section)
2. MgB$_2$ ceramics (a scale in cm)
3. X-ray diffraction pattern of MgB$_2$ powder
4. AC-susceptibility of MgB$_2$ ceramics
5. Resistance of 1400°-annealed MgB$_2$ ceramics as function of temperature

Fig. 1

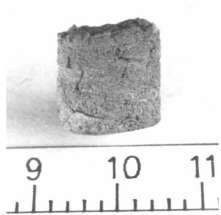

Fig. 2

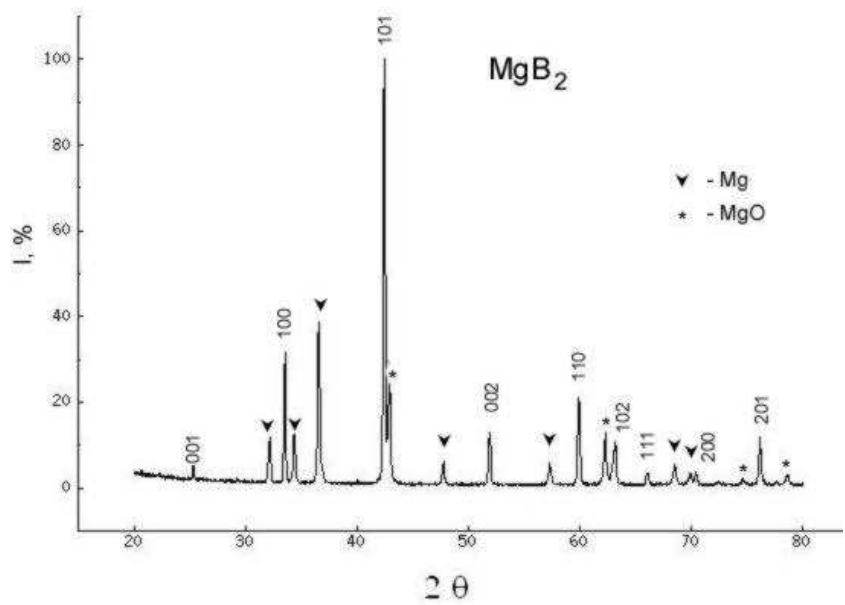

Fig. 3

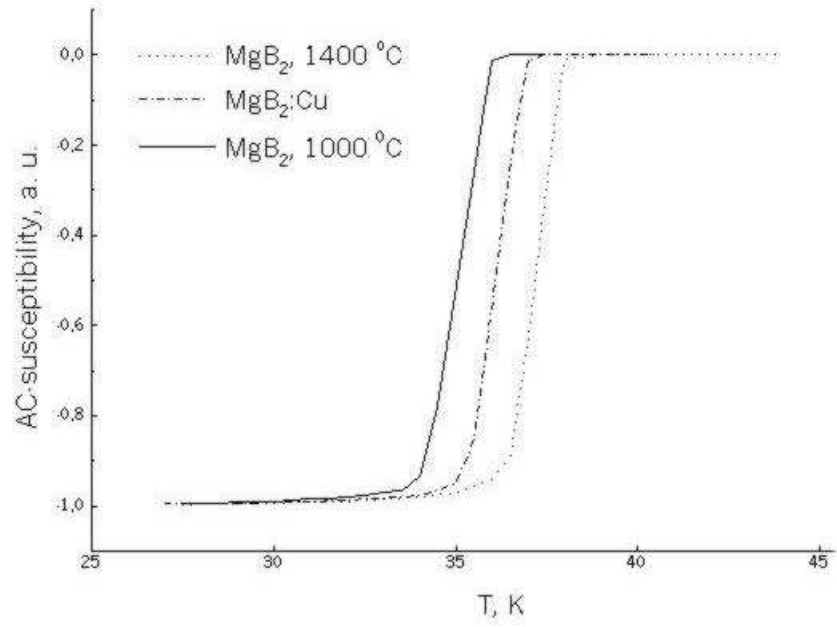

Fig. 4

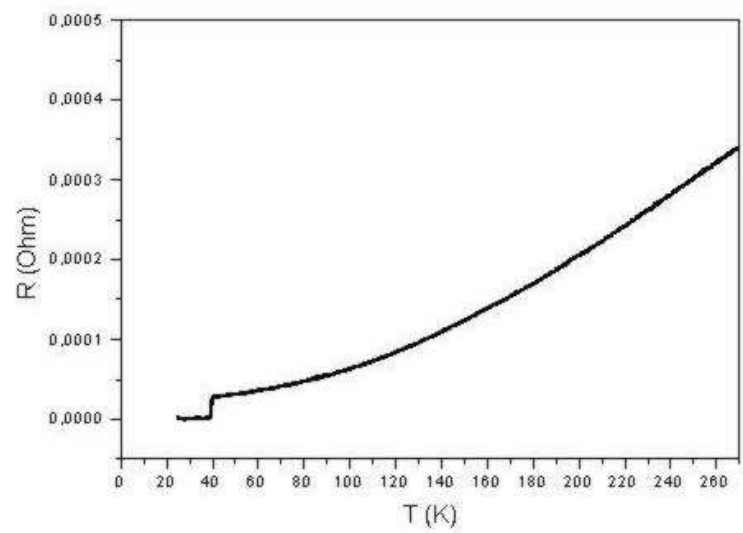

Fig. 5